\documentstyle[rmaaconf]{article}
\begin{document}

{\it to be published in the ``Starburst Activity in Galaxies''  
proceedings by the RevMexAstronAstrofis. (ConfSeries)}

\title{UV HST Spectroscopy of Star-Forming Galaxies}

\author{D. Kunth\altaffilmark{1}, J. Lequeux\altaffilmark{2},  J. M.
 Mas-Hesse\altaffilmark{3}, E. Terlevich\altaffilmark{4} \& R.
 Terlevich\altaffilmark{5}}
\altaffiltext{1}{Institut d'Astrophysique de Paris, 98 bis Bd. Arago,
 F-75014 Paris, France}
\altaffiltext{2}{DEMIRM, Observatoire de Paris}
\altaffiltext{3}{LAEFF-INTA, POB 50727, E-28080 Madrid, Spain}
\altaffiltext{4}{Institute of Astronomy: University of Cambridge, UK}
\altaffiltext{5}{Royal Greenwich Observatory}
\begin{resumen}
Informamos sobre las observaciones espectrosc\'opicas  de 8 galaxias HII, realizadas con el HST. Se ha detectado emisi\'on en la l\'{\i}nea Ly$\alpha$ 
en 4 de ellas. Hemos encontrado que el factor determinante para que la 
l\'{\i}nea sea visible es la estructura en velocidades del gas neutro, y no la 
abundancia de polvo. El resto de las galaxias observadas muestran l\'{\i}neas 
de absorci\'on en Ly$\alpha$ anchas y saturadas, atribuidas a densidades de columna altas de HI est\'atico con relaci\'on a los fotones Ly$\alpha$ emitidos 
en las regiones HII. Las galaxias con formaci\'on estelar IZW18 y, 
especialmente, SBS0335--052 pudieran albergar nubes de HI extremadamente 
deficientes en metales, con un valor de [O/H] para \'esta \'ultima tan bajo 
como -7.2.
\end{resumen}

\begin{abstract}
HST spectroscopical observations of 8 HII galaxies are reported. Ly$\alpha$
emission was detected in 4 of them. We find that it is the velocity
structure of the gas which is the main determining factor for the escape of the
Ly$\alpha$ photons, and not the abundance of dust. The rest of the sample
shows broad damped Ly$\alpha$ absorption attributed to large HI column
densities that is static with respect to the emitted Ly$\alpha$ photons
emerging from the HII regions. The star-forming galaxies IZW18 and even
more SBS0335--052 may have extremely metal deficient HI clouds, the latter
with [O/H] as low as -7.2.
\end{abstract}


\section{Introduction}
  
A very important astrophysical issue is the detection of galaxies at large
redshift that are forming stars for the first time, the so-called primeval
galaxies. In parallel, bearing in mind that galaxy formation may not be
assigned to any preferential cosmological epoch but rather a continuous
process, one might find left-over pristine gas pockets that are
forming young galaxies at present epoch. For this reason, since in our
local universe there may be star-forming galaxies that look like very much
distant primeval ones, there have been several attemps to observe their
Ly$\alpha$ emission (Meier and Terlevich  1981).

Studies have also been aimed to measure abundances in the neutral gas of
gas--rich dwarf galaxies with spectra dominated by recent star formation
episodes.  Indeed, in objects such as these, the HI clouds largely extend beyond
the optical images suggesting that a substantial fraction of this gas might
still be chemically unevolved or even pristine (Roy and Kunth 1995).  With
the advent of the HST, it became possible to analyse with much greater
details than with the IUE the processes of escape and the destruction of
the Ly$\alpha$ photons since a study of the line profile could be
performed.  Similarly at a spectral resolution of 20000 it became possible
to disentangle nebular from stellar absorption lines and give estimates of
the metal abundances in the insterstellar medium.

\section{The IUE era}

Previous IUE observations were performed on more than a dozen galaxies in
the SWP low resolution mode (Meier and Terlevich 1981; Hartmann, Huchra and
Geller 1984; Deharveng, Joubert and Kunth 1986; Hartmann et al. 1988 and
Terlevich et al. 1993).  Galaxies with redshift large enough that their
Ly$\alpha$ emission is separated from the geocoronal line were selected. It
was realized from the very beginning that the Ly$\alpha$ emission is much
weaker, by an order of magnitude, than expected from the recombination
theory. In fact the equivalent widths of this line are in the range of 
50 to 10~\AA\ or
lower and moreover in several cases it is seen in absorption. No obvious ways
were found to predict the outcome in a general case. Previous works have
also shown a possible anticorrelation between the Ly$\alpha$/H$\beta$ ratio
and the metallicity (actually the O/H abundance, as measured in the ionised
gas).  These results and the lack of ``primeval galaxies" at large redshift
in blank sky searches for redshifted Ly$\alpha$ emission has been
attributed to the effects of dust absorption that preferentially destroys
Ly$\alpha$ photons (Charlot and Fall 1993, and references therein). The
process behind is that the transfer of Ly$\alpha$ radiation is strongly
affected by resonant scattering from neutral interstellar hydrogen
atoms. By increasing enormously their path length, photons become more
vulnerable to dust absorption. Chen and Neufeld (1994) have shown that the
combination of interstellar dust absorption and hydrogen atoms scattering
can even lead to negative Ly$\alpha$ equivalent widths as observed in IZW18
by Kunth et al. (1994) using HST data.  Since IZW18 is  the most
metal-poor galaxy known it soon became clear that the transport of
Ly$\alpha$ photons may not be attributed to the galaxy dust content
alone. On the other hand a positive emission has been detected in the more
dusty galaxy Haro2, (Lequeux et al. 1995). These new facts and the new
capability of the HST to analyse in higher details for the first time
Ly$\alpha$ line profiles in nearby galaxies led us to embark on a longer
term project using the GHRS.

\section{The new HST observations}

Observations were made using the same settings as in Kunth et al. (1994)
and Lequeux et al. (1995), the grating angle being set according to the
redshift of the object, so as to cover the Ly$\alpha$ and the OI
1302.2~\AA\ regions respectively. The Ly$\alpha$ range was chosen to
investigate the Ly$\alpha$ photon escape and measure the column density of
the surrounding neutral gas on the line of sight. The OI 1302~\AA\ and SiII
1304~\AA\ region was used to estimate the chemical composition of the gas
and to measure with reasonable accuracy the mean velocity at which the
absorbing material lies with respect to the star-forming region of a given
galaxy. Eight galaxies have been observed so far, as listed in
Table~\ref{tab:galaxies}, and have been selected from very different
considerations:

\begin{table}[t]
  \caption{Observed HII galaxies} 
  \label{tab:galaxies}
  \begin{center}\leavevmode
    \begin{tabular}{ccccc}\hline
      Galaxies & $\alpha$1950 & $\delta$1950 & v(km/sec) & Ly$\alpha$ \\ 
      \hline
      ESO 350-IG038&00h34m26.0&-33d49m54&6156&emission \\
      SBS0335-052& 03h35m15.1&-05d12m27&4043& \\ 
      IRAS08339+6517&08h33m57.3&+65d17m45&5730&emission \\
      IZW18 &09h30m30.3&+55d27m46& 740& \\
      Haro 2 &10h29m22.7&+54d39m31&1461&emission \\
      Mkn36 &11h02m15.6&+29d24m28& 646& \\
      IIZW70  & 14h48m55.1&+35d46m37&1215& \\
      ESO 400-G043&20h34m31.0&-35d39m42&5900&emission \\ \hline
    \end{tabular}
  \end{center}
\end{table}

i) the HII galaxies Mkn36, IIZW70 and Haro2 were chosen because
they span a wide range of metallicity. The aim was to investigate the
possible relationship between the composition of their HII regions and that
of the HI gas as derived from the OI and SiII lines.

ii) Three starburst galaxies were selected in the IUE-ULDA from the
a-priori knowledge that they were Ly$\alpha$ emitters; they include: IRAS
08339+651, ESO 350-IG038 and ESO 400-G043.

In addition the SBS0335-052 spectra, observed by  Thuan, Isotov, and
Lipovetsky  were retrieved from the HST archives.

\section{THE LYMAN ALPHA ESCAPE}

Among the eight galaxies up to now observed with the GHRS, 4 show no
Ly$\alpha$ emission.  Instead, a strong damped Ly$\alpha$ absorption
redshifted at the rest velocity is observed, showing a complete destruction
of these photons in the nebular gas.  OI and/or SiII appear in absorption
and in some cases (see section 5) are barely detected. In all cases these
lines occur without any velocity shift with respect to the HII
regions. This indicates that the neutral gas in which they mostly originate
is static with respect to the star-forming region. Therefore, although
these galaxies are relatively dust free (IZW18 shows weak signs of reddening
and its gas-to-dust ratio is at least 50 times larger than the Galactic value
- Kunth et al. 1994) it is possible to suppress Ly$\alpha$ by simple
multiple resonant scattering from the neutral gas. Even a dust free
cloud would nearly preserve the photons but scatter them over the whole HI
cloud area; the surface brightness in the line would then be very weak but
might perhaps be detected with the HST with very long exposures.
In all these cases the widths of the
broad damped absorptions are larger than 20~\AA\ (33~\AA\ in the case of
IZW18) hence  one can dismiss the possibility that they arise in OB 
photospheres (Valls-Gabaud, 1993). 

At variance with these cases, the rest of the sample shows clear Ly$\alpha$
emission. The profiles of the lines are nevertheless asymmetric, with the
peak emission REDSHIFTED with respect to the HII region systemic velocity.

The first bonafide case was reported in Haro 2 in which the Ly$\alpha$
emission is accompanied by a broad absorption in the blue wing of that line
(Lequeux et al. 1995), with the general appearance of a typical P Cyg
profile. Therefore, the neutral gas responsible for the
absorption in this galaxy is not at the velocity of escaping photons.
In other words multiple scattering at wavelengths larger than the Ly$\alpha$ 
rest wavelength is not effective, so that Ly$\alpha$ photons can escape in the
red wing. This may be due to an expanding envelope around the HII region,
whose back part emits or scatters Ly$\alpha$ photons which are not absorbed
by the front part because of its different velocity.
This interpretation is confirmed by the presence of other
detected absorptions of OI, SiII and SiIII due to gas in front of the hot
stars ionizing the central HII region of Haro 2. The heliocentric
velocities of all these absorptions are lower by about 200 km/sec than that
of the bulk of the galaxy as measured in the 21-cm line and of the optical
emission lines. Moreover, the absorption at the rest wavelengths were
almost negligible for all metallic lines. 

Spectroscopic observations of H$\alpha$ were obtained with the William 
Herschell
telescope at La Palma, allowing to compare both the reconstructed
Ly$\alpha$ and the H$\alpha$ profiles. Prelimirary results show that
significantly Ly$\alpha$ is broader than the Balmer line suggesting the
scattering of the photons from the back side of the expanding neutral
cloud. The amount of neutral gas that produces the blue absorption trough at
Ly$\alpha$ is rather modest and of the order of N(HI)=7.7~10$^{19}$
atom/cm$^{2}$.

\begin{figure}
\vspace*{75mm}
\includegraphics{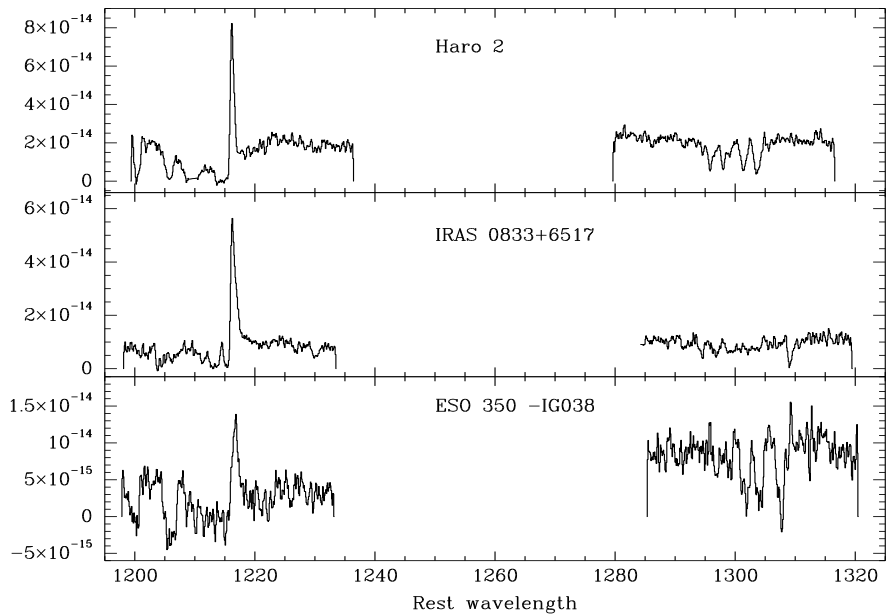} 

\caption{HST-GHRS spectra for 3 galaxies with Ly$\alpha$ emission and showing
an absorption trough. Both the  Ly$\alpha$ and the OI regions are displayed.}
\label{fig:haro}
\end{figure}

The three additional galaxies that have since been observed by us with the
HST using the GHRS remarkably confirm that Haro 2 is not an isolated
case. We display these spectra in  Fig.~\ref{fig:haro}.

All spectra show Ly$\alpha$ emission with a broad absorption on their blue
side. Careful analysis  to interpret the details of their
profiles is being published elsewhere. The preliminary results are as follows: 

\begin{itemize} 

\item ESO 350-IG038: This galaxy has a redshift of 6156 km/sec. At this
redshift the OI region falls close to the CII 1326~\AA\ Galactic line that
can be used as a reference for the wavelength scale. The Ly$\alpha$
absorption extends over more than 1500 km/sec to the blue side of the line
emission. It is not clear that Ly$\alpha$ exhibits a P-Cygni profile
since the line does not drop sharply at zero velocity. The OI and SiII 
lines are very broad and peak at -80km/sec with respect to the HII gas.

\item IRAS 0833+6517: This galaxy has a redshift of 5730 km/sec. The Ly$\alpha$
is narrow and exhibits a clear P Cygni profile. Remarkable enough is a
clear secondary emission, probably from Ly$\alpha$ at -200 km/sec from the
main line.  The absorption feature 400 km/sec wide  is clearly seen
on the blue side of the profile. Unfortunately no OI and/or SiII are
detected that could provide more detailed information about the kinematics
of the absorbing gas. 

\item  ESO-B400: This galaxy was observed soon after this Meeting and the
results are presented elsewhere. Again
absorptions are detected and take place in flowing out material. \\

\end{itemize}

The main conclusions that are drawn  from this set of data is that
complex velocity structures are dominant in the Ly$\alpha$ emission,
showing the strong energetic impact of the star-forming regions into
their surrounding ISM. {\it It is this velocity structure the
determining factor for the Ly$\alpha$ escape, not the abundance of 
dust}. This effect helps to understand why only luminous high-redshift
objects have been found up to now with linewidths larger than 1000 km/sec.
High-redshift galaxies with very strong (EWs $>$ 500~\AA\ ) extended
Ly$\alpha$ emission are characterized by strong velocity shear and
turbulence (v $>$ 1000 km/sec) and this suggests that other ionization
mechanism than photoionization by young stars may be operating. However
Steidel et al. (1996) have recently discovered a substantial population of
star--forming galaxies at 3.0$<$z$<$3.5 that were selected not from their
emission--line properties but from the presence of a very blue far-UV
continuum and a break below 912~\AA\ at rest frame. Similarly to our local
starbursts they find that 50\% of their objects show NO Ly$\alpha$ emission
whereas the rest does, but with weak EWs no larger than 20~\AA\ at rest.
This population looks indeed very similar to our local starburst galaxies.

\section{MEASURING NEUTRAL ABUNDANCES}

\begin{figure}
\vspace*{75mm}

\includegraphics{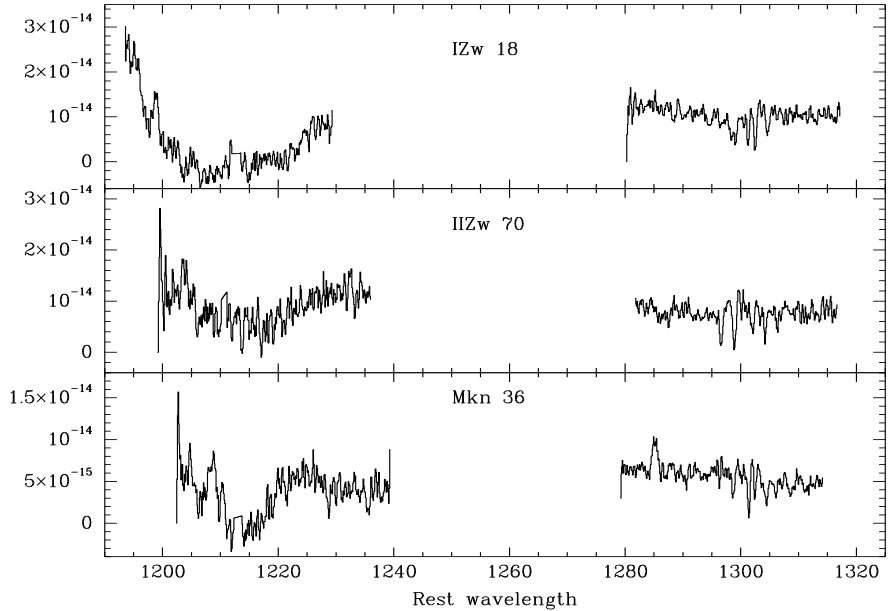} 

\caption{HST-GHRS spectra for 3 galaxies with no Ly$\alpha$ emission. In this
case the Ly$\alpha$ region shows a broad damped absorption. Both the
  Ly$\alpha$ and the OI regions are displayed.}
\label{fig:mkn}
\end{figure}

Since blue compact galaxies are rich in neutral gas which might remain
unprocessed they were thought to be ideal laboratories to look for
primordial gas. Alternatively, because they undergo sporadic episodes of
massive star-formation it is expected that their ISM remains inhomogeneous.
One could test the mechanisms by which metals are dispersed and further
mixed into the ISM (Kunth and Sargent 1986; Roy and Kunth 1995;
Tenorio-Tagle 1996). Neutral heavy elements abundance 
informs about  past star-formation episodes after mixing has been operating. 
 It is remarkable that whenever  Ly$\alpha$ emission is not
detected  broad damped absorption features are detected. As noted before
the lines are too wide to be attributed to stellar photospheres. In IZw 18,
Mkn 36 and IIZw 70 the OI and SiII lines were detected or well measured. 
IZw 18 has
been analysed in Kunth et al. (1994) who find N(HI)=3.5~10$^{21}$
atom/cm$^{2}$ on the line of sight within the LSA aperture of 2"x2". The
authors concluded that most of the OI is produced in the HI gas and very
little in the transition zone of the HII gas.  Accordingly they conclude
that the O/H abundance in the HI region is a factor of about 20 BELOW that
in the HII region. Although words of caution were given as regarding the
uncertainties involved with the analysis of the OI 1302~\AA\ line, the
result indicates that most of the heavy elements have been produced in the
present burst of star formation (see also Kunth el al. 1995). Pettini and
Lipman (1995) have added some illustrative arguments to moderate the impact
of this result: indeed if OI is saturated (this point is not completely
settled) the OI profile is more sensitive to the b-value than to the column
density.  Additional HST observations are scheduled to solve this question
using the SII 1256~\AA\ line, that is expected to remain unsaturated and is
mostly produced in the HI gas.  This result prompted us to investigate more
galaxies with the hope to correlate HI O abundances with that of the HII
zone. As can be seen in Fig.~\ref{fig:mkn} absorptions in Mkn 36 and IIZw 70 are not as broad as in
IZw 18.  Nethertheless they indicate very large N(HI) column densities.

OI equivalent widths of the order of 0.3 \AA\, together with OI and HI column
densities estimated with the XVoigt code, lead to O abundances similar to
that of the HII regions.

\begin{figure}
\vspace*{45mm}
\includegraphics{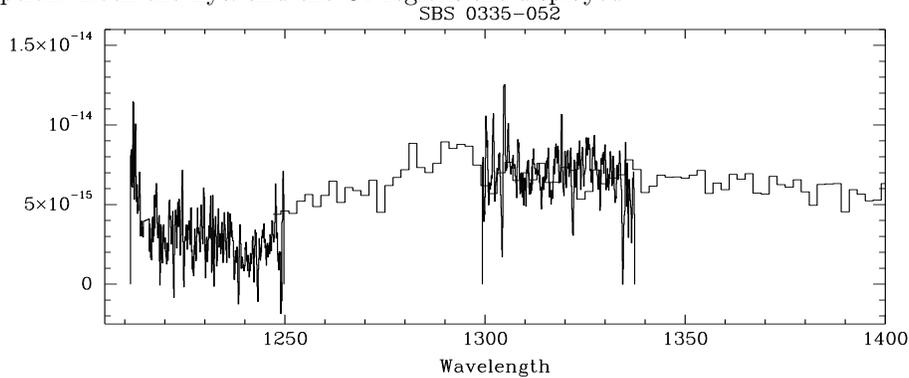} 

\caption{HST-GHRS  spectra of SBS 03350052 for the
 Ly$\alpha$ and the OI regions superimposed to the IUE SWP spectrum.}
\label{fig:sbs}
\end{figure}

The case of SBS 0335-052 is rather different. We have de-archived the HST
GHRS spectra that were obtained by Thuan, Isotov, and
Lipovetsky using the same settings as in
Kunth et al. (1994) for IZw 18. The spectrum is much noiser
than the rest of the sample. Nethertheless we made use of the IUE spectrum
that had been obtained by Terlevich E. and Terlevich R. using a
 combined
NASA-ESA shift (de-archived). Fig.~\ref{fig:sbs} shows
the HST spectra superimposed - after proper scaling - to the IUE SWP
spectrum near the Ly$\alpha$ and OI regions and displayed on the observed
wavelength scale.

 We have barely been able to
detect the OI line at 1302\AA\ . Using the red wing of the Ly$\alpha$
absorption we were able to fit a damped line with a b value of around 40
km/sec leading to a columm density N(HI)=3.5~10$^{21}$ atom/cm$^{2}$, even
higher than that of IZW18.  This combined information leads to logN(OI) of
14.3 or to a [O/HI] ratio as low as -7::! If real this would indicate that
SBS0335-052 has really undergone very little star formation - if any - in
the past. This makes this target ideal for primordial He determination.

Note that these determinations are very preliminary. It can be seen in
Fig.~\ref{fig:sbs} that the bottom of the Ly$\alpha$ absorption does not
 reach the zero
level as it should. We suspect that this is due to the extraction procedure
which may lack some accuracy at low level of background subtraction
(ie. $<$ 2x10$^{-15}$ erg/sec/cm$^{2}$/\AA\ ). The summary of the abundance
measurements are given in Table~\ref{tab:abundance}

\begin{table}[t]
  \caption{Abundances in HI gas} 
  \label{tab:abundance}
  \begin{center}\leavevmode
    \begin{tabular}{ccccc}\hline
 
      Galaxies & $LogN(HI)$ & $LogN(OI)$ & $[O/HI]$ & $HI/HII abund ratio$ \\ \hline
      IZW18 & 21.5 & 15.4 & -6.1:: & $\ll$1  \\
      IIZW70  & 20.6 & 14.9 & -5.7:: & 1  \\
      Mkn36 & 19.9 & 14.4 & -5.5:: & 1  \\
      SBS0335-052& 21.5 & 14.3 & -7.2:: & $\ll$1  \\ \hline
    \end{tabular}
  \end{center}
\end{table}

\section{SUMMARY}

\begin{itemize} 

\item Ly$\alpha$ emission has been observed in 4 HII galaxies out of 8
observed with the GHRS onboard HST.  We have found that the determining
factor to allow the escape of Ly$\alpha$ photons is the velocity structure
of the neutral gas (and may be the presence of holes with low column
densities), and not the abundance of dust particles. In fact one galaxy in
our sample is a strong IRAS source (IRAS 0833+6517).  Whenever the HI
column density is large enough, even in a dust free environnement (IZw 18 is
the best example) photons can be completely re-distributed by multiple
scattering, presumably over the area of the associated HI clouds. The
Ly$\alpha$ line then becomes very hard to detect because of its low surface
brightness.  The
photons so redistributed will correspond to the Ly$\alpha$ photons emitted
by the HII region only if the HI gas is static with respect to the HII gas. 

\item A clear evidence for the presence of a wide velocity field is
given by the presence of a deep absorption trough in the blue side of the
Ly$\alpha$ profile. Moreover, absorption lines of metallic elements (OI,
SiII) are also detected significantly blueshifted with respect to the HII
gas velocity. This outflowing neutral material may eventually leave the
galaxy. We thus may be witnessing galactic wind that results from intense
star-forming activity.

\item Several possibilities are investigated to understand the reasons
that govern the appearance of the Ly$\alpha$ line emission. The age of the
burst, its strength, the metallicity of the gas (controlling the cooling,
hence the blow out time occurence), the gravitational potential of the
parent galaxy and its morphology, the HI and the dust distributions could
play a role but of unequal importance that we hope to assess in a near
future.

\end{itemize}

\end{document}